\documentclass[aps,nofootinbib]{revtex4}

\usepackage{epsfig}
\usepackage{latexsym}
\usepackage{graphicx}
\usepackage{amsmath}
\usepackage{amsfonts}   
\usepackage{amssymb}    
\usepackage{subfigure}

\newcommand{\newc}{\newcommand}

\newc\eg{{\rm {e.g.}}}  \newc\etal{{\rm {et al.}}} \newc\ie{{\rm i.e.}}
\newc\etc{{\rm {etc}}}

\newcommand\lsim{\mathrel{\rlap{\lower4pt\hbox{\hskip1pt$\sim$}}
    \raise1pt\hbox{$<$}}}
\newcommand\gsim{\mathrel{\rlap{\lower4pt\hbox{\hskip1pt$\sim$}}
    \raise1pt\hbox{$>$}}}

\newc{\egamma}{E_{\gamma}}
\newc{\flux}[1]{\Phi_{#1}}
\newc{\dfluxde}[1]{\frac{d\Phi_{#1}}{d E_{#1}}}

\newc{\fluxg}[1]{\Phi_{\gamma}({#1})}
\newc{\fluxggc}{\Phi_{\gamma}^{\rm GC}} \newc{\fluxgegb}{\Phi_{\gamma}^{\rm EGB}}
\newc{\rfluxggc}{R_{\Phi_\gamma}^{\rm GC}}

\newc{\dfluxgde}{\frac{d\Phi_{\gamma}}{d\egamma}}
\newc{\dfluxgdetext}{ d\Phi_{\gamma} / d\egamma}
\newc{\rdifffluxggc}{R_{\dfluxgdetext}^{\rm GC}}

\newc{\intensityg}{I_{\gamma}}

\newc{\jpsi}{J(\psi)}
\newc{\jpsiave}{\langle \jpsi \rangle} 
\newc{\jpsizeroave}{\langle J(\psi=0) \rangle} 
\newc{\javearea}{\langle J \rangle} 
\newc{\deltaomega}{{\Delta\Omega}}

\newc{\eplus}{e^+}
\newc{\epos}{E_{\eplus}}
\newc{\eps}{\varepsilon}

\newc{\npos}{n_{\eplus}} \newc{\Npos}{N_{\eplus}}
\newc{\dnposde}{\frac{d n_{\eplus}}{d\epos}}
\newc{\dnposdeps}{\frac{d n_{\eplus}}{d\eps\phantom{_{\eplus}}}}
\newc{\dnposdepstext}{ d n_{\eplus} / d\eps}
\newc{\fluxpos}{\Phi_{\eplus}}  \newc{\fluxelec}{\Phi_{e^{-}}}
\newc{\dfluxposde}{\frac{d\Phi_{\eplus}}{d\epos}}
\newc{\dfluxposdetext}{ d\Phi_{\eplus} / d\epos}

\newc{\mhalf}{m_{1/2}}      \newc{\mzero}{m_0}
\newc{\tanb}{\tan\beta}
\newc{\azero}{A_0}
\newc{\at}{A_t} \newc{\ab}{A_b} \newc{\atau}{A_\tau}
\newc{\bmu}{B\mu}           \newc{\sgn}{{\rm sgn}}
\newc{\mone}{M_1}           \newc{\mtwo}{M_2}

\newc{\charone}{\chi_1^\pm} \newc{\mcharone}{m_{\chi_1^\pm}}

\newc{\hl}{h}               \newc{\mhl}{m_{\hl}}   \newc{\gammahl}{\Gamma_{\hl}}
\newc{\hh}{H}               \newc{\mhh}{m_{\hh}}   \newc{\gammahh}{\Gamma_{\hh}}
\newc{\ha}{A}               \newc{\mha}{m_{\ha}}   \newc{\gammaha}{\Gamma_{\ha}}
\newc{\hpm}{H^{\pm}}        \newc{\mhpm}{m_{\hpm}} \newc{\gammahpm}{\Gamma_{\hpm}}
\newc{\hp}{H^{+}} \newc{\mhp}{m_{\hp}} \newc{\hm}{H^{-}}
\newc{\mhm}{m_{\hm}}
\newc{\xt}{X_{t}}           \newc{\xb}{X_{b}}

\newc{\amu}{a_{\mu}}        \newc{\amususy}{a_{\mu}^{\text{SUSY}}}
\newc{\amuexpt}{a_{\mu}^{\text{expt}}}        \newc{\amusm}{a_{\mu}^{\text{SM}}}
\newc{\deltaamususy}{\delta a_{\mu}^{\text{SUSY}}}
\newc\gmtwo{(g-2)_{\mu}} \newc\deltaamu{\Delta a_{\mu}}
\newc{\msbar}{\overline{MS}} \newc{\drbar}{\overline{DR}}
\newc{\yt}{h_t} \newc{\yb}{h_b} \newc{\ytau}{h_{\tau}}

\newc{\mtop}{m_t}               \newc{\mtpole}{M_t}
\newc{\mtaupole}{m_{\tau}^{\text{pole}}}
\newc{\mtmtsmmsbar}{m_t(m_t)^{\msbar}_{{\text{SM}}}}
\newc{\mtmtsmdrbar}{m_t(m_t)^{\drbar}_{{\text{SM}}}}
\newc{\mtmtmssmdrbar}{m_t(m_t)^{\drbar}_{{\text{SUSY}}}}

\newc{\mbmbmsbar}{m_b(m_b)^{\msbar} }

\newc{\mbmbsmmsbar}{m_b(m_b)^{\msbar}_{{\text{SM}}}}
\newc{\mbmzsmmsbar}{m_b(\mz)^{\msbar}_{{\text{SM}}}}
\newc{\mbmzsmdrbar}{m_b(\mz)^{\drbar}_{{\text{SM}}}}
\newc{\mbmzmssmdrbar}{m_b(\mz)^{\drbar}_{{\text{SUSY}}}}

\newc{\mtaumzsmmsbar}{m_{\tau}(\mz)^{\msbar}_{{\text{SM}}}}
\newc{\mtaumzsmdrbar}{m_{\tau}(\mz)^{\drbar}_{{\text{SM}}}}
\newc{\mtaumzmssmdrbar}{m_{\tau}(\mz)^{\drbar}_{{\text{SUSY}}}}

\newc{\mgut}{M_{\rm GUT}}
\newc{\mplanck}{M_{\rm P}}      \newc{\mpl}{M_{\text{Pl}}}
\newc{\msusy}{M_{\rm SUSY}}      \newc{\ms}{M_{\text{S}}}
\newc{\jxf}{J({\xf})}
\newc{\jxfexact}{J_{\rm exact}({\xf})}  \newc{\jxfexp}{J_{\rm exp}({\xf})}
\newc{\VEV}[1]{\langle #1 \rangle}
\newc{\xf}{x_f}
\newc\vrel{v_{\rm rel}}

\newcommand\mchi{m_{\chi}}

\newc\sell{{\widetilde e}_L}      \newc\msell{m_{\sell}}
\newc\selr{{\widetilde e}_R}      \newc\mselr{m_{\selr}}
\newc\snue{{\widetilde \nu}_e}      \newc\msnue{m_{\snue}}
\newc\snutau{{\widetilde \nu}_\tau}      \newc\msnutau{m_{\snutau}}
\newc\supl{{\widetilde u}_L}      \newc\msupl{m_{\supl}}
\newc\supr{{\widetilde u}_R}      \newc\msupr{m_{\supr}}
\newc\sdl{{\widetilde d}_L}      \newc\msdl{m_{\sdl}}
\newc\sdr{{\widetilde d}_R}      \newc\msdr{m_{\sdr}}

\newcommand\squark{\widetilde q}
    \newcommand\msquark{\widetilde{m}_{\squark}}

\newc\stoponetwo{{\widetilde t}_{1,2}}
\newc\sbotonetwo{{\widetilde b}_{1,2}}
\newc\stauonetwo{{\widetilde \tau}_{1,2}}

\newc\sfermion{\tilde f}  \newc\msfermion{m_{\sfermion}}

\newc{\gstar}{g_\ast}           \newc{\gsstar}{g_{s\ast}}
\newc{\geff}{g_{\rm eff}}
\newcommand\mz{m_{Z}}

\newc{\sthw}{\sin\theta_W}              \newc{\cthw}{\cos\theta_W}
\newc{\bino}{\widetilde B}              \newc{\wino}{\widetilde W_30}
\newc{\higgsinob}{{\widetilde H}^0_b}   \newc{\higgsinot}{{\widetilde H}^0_t}

\newc{\sigmav}{\sigma v}

\newc{\abund}{\Omega h^2}
\newc{\abundchi}{\Omega_\chi h^2}
\newc{\abundcdm}{\Omega_{\text{CDM}} h^2}

\newc{\omegachi}{\Omega_\chi}
\newc{\omegam}{\Omega_{M}}       \newc{\abundm}{\Omega_{M} h^2}
\newc{\omegab}{\Omega_{b}}       \newc{\abundb}{\Omega_{b} h^2}
\newc{\omegacdm}{\Omega_{CDM}}
\newc{\omegatot}{\Omega_{TOT}}
\newc{\rhocrit}{\rho_{crit}}
\newc{\rhochi}{\rho_{\chi}}

\newcommand{\mev}{\mbox{ MeV}}
\newcommand{\gev}{\mbox{ GeV}}
\newcommand{\tev}{\mbox{ TeV}}

\newc\cmeter{{\rm cm}} \newc\meter{{\rm m}} \newc\kmeter{{\rm km}}

\newc\second{{\rm s}} 
\newc\sr{{\rm sr}}

\newc\pb{\,\mbox{pb}} \newc\fb{\,\mbox{fb}}

\newc\pc{\,\mbox{pc}} \newc\kpc{\,\mbox{kpc}}
\newc\mpc{\,\mbox{Mpc}} \newc\gpc{\,\mbox{Gpc}}

\newc\BR{BR}
\newc\bsgamma{b\rightarrow s \gamma }
\newc\bxsgamma{\overline{B}\rightarrow X_{s}\gamma}
\newc\brbsgamma{\BR(\overline{B}\rightarrow X_s\gamma)}

\newc{\beq}{\begin{equation}}
\newc{\eeq}{\end{equation}}
\newc{\bea}{\begin{eqnarray}}
\newc{\eea}{\end{eqnarray}}
\newc\eqn[1]{Eq.~(\ref{#1})}

\newc{\sigsip}{\sigma^{SI}_{p}} \newc{\sigsin}{\sigma^{SI}_{n}}
\newc{\sigsiN}{\sigma^{SI}_{N}}
\newc{\sigsdp}{\sigma^{SD}_{p}} \newc{\sigsdn}{\sigma^{SD}_{n}}
\newc{\sigsiA}{\sigma^{SI}_{A}}

\newc{\pbar}{\bar{p}}

\newc{\chisq}{\chi^2}  \newc{\chisqred}{\chi^2_{\text{red}}}

\newc{\nfwc}{{\text{NFW+ac}}} \newc{\moorec}{{\text{Moore+ac}}}

\newc\xilim{\xi_{\rm lim}} 
\newc\tlim{t_{\rm lim}} 
\newc\zetalim{\zeta_{\rm lim}} 

\newc\zetah{\zeta_h}
\newc{\relprobone}[1]{p({#1} \vert d)}
\newc{\relprobtwo}[2]{p({#1},{#2} \vert d)}

\long\def\begincomment#1\endcomment{%
        \begingroup\sf\baselineskip12pt#1\endgroup}

\newcommand\app[3]   { 
		{{Astropart. Phys.\ }{\bf #1} (#2) #3}}
		
\newcommand\prep[3]   { 
		{{Phys. Rep.\ }{\bf #1} (#2) #3}}				
			
\newcommand\aphj[3]   { 
		{{Astrophys. J.\ }{\bf #1} (#2) #3}}

\begin{document}
\title{Shedding Light on Dark Matter with Fermi LAT Data on Gamma Rays}
\author{Leszek Roszkowski$^{1,2}$ and Yue-Lin Sming Tsai$^{1}$}
\affiliation{
$^1$Department of Physics and Astronomy, University of Sheffield,
Sheffield S3 7RH, England \\
$^2$The Andrzej
Soltan Institute for Nuclear Studies, Warsaw, Poland\\
}


\begin{abstract}
  The diffuse Galactic $\gamma$--ray data from the region of the
  Galactic Center has been collected by the LAT instrument on the
  Fermi Gamma-Ray Space Telescope. In this paper we argue that it may
  be able to provide an unambiguous evidence of originating, in
  addition to known astrophysical sources, from dark matter
  annihilations in the halo, independently of the mass and other
  properties of the dark matter particle. We also show that the
  recently released high precision data from mid-latitudes is already
  providing an upper bound, albeit still a weak one, on the cuspiness
  of the dark matter density profile as a function of the mass of the
  dark matter assumed to be a stable neutralino of minimal supersymmetry.
\end{abstract}


\date{\today}

\maketitle



\section{Introduction}
The evidence for the existence of dark matter (DM) in the Universe is
strong and mounting but its nature remains unknown. It is generally
believed that it is made up of some unknown weakly interacting massive
particle (WIMP) for which there are a whole host of possible
hypothetical candidates predicted in various extensions of the Standard
Model of particle physics, with the lightest neutralino of
supersymmetry (SUSY) being perhaps the most popular
one~\cite{susy-dm-reviews}.  Nevertheless, it is clear that the mass
and other properties of DM can only be established experimentally. A
whole range of detection strategies have been developed to search for
DM. A number of underground detectors are currently in operation
trying to directly detect a faint signal from a Galactic halo WIMP
scattering off a target and, if successful, they would provide perhaps
the most unambiguous signal of DM.  Indirect searches look for exotic
products of WIMP annihilation (or decay, if unstable) in the Sun's or
Earth's core, in the Galactic halo or the Galactic Center
(GC)~\cite{susy-dm-reviews}. Finally, WIMP  particles are likely to be produced at the LHC.

Following last year's launch of the Fermi Gamma-Ray Space Telescope
(Fermi), $\gamma$--ray window on the sky has received a major boost
with new unprecedented quality data on pulsars~\cite{Abdo:2008nz},
gamma-ray-bursters~\cite{Abdo:2009zz}, diffuse radiation, \etc. The
diffuse radiation in particular is of much interest since, under
favorable conditions, it may reveal a measurable contribution from
annihilation products of dark matter~\cite{susy-dm-reviews}. Such
favorable conditions may exist in the region of the GC where
DM density is believed to be enhanced, and Fermi data from that region
are eagerly awaited by the dark matter community. In the meantime, high precision
data from Fermi LAT for diffuse emission with photon energies of
$0.1\gev$ to $10\gev$ from the region of intermediate Galactic
latitudes $10^{\circ}<|b|<20^{\circ}$ and the full range of the
Galactic longitudes ($0\leq l<360^{\circ}$) has recently been
presented~\cite{fermimidl,porter_lodz09}.

In this paper we make two points. The first is that, by comparing
upcoming Fermi data on $\gamma$--ray flux at different angular
distances from the Galactic Center, one may be able to unambiguously
infer its DM origin, independently of the dark mater mass and other
properties. The second point is that, the recently released Fermi data
for mid-latitude $\gamma$--rays already provides a constraint on the
profile of the Galactic DM halo as a function of DM mass assumed to be
the lightest neutralino of supersymmetry.

The paper is organized as follows. In Sec.~\ref{sec:grdm} we briefly
review the formalism used to compute $\gamma$--rays from DM
annihilation and list some popular DM halo profiles used subsequently
in Sec.~\ref{sec:dmtestgc} to suggest tests of DM signatures in the GC
and to derive, in Sec.~\ref{sec:midlatconstr}, an upper bound on the
cuspiness of the DM halo profile. In Sec.~\ref{sec:summary} we briefly
summarize our results.

\section{Diffuse Gamma-rays from Galactic dark matter annihilation}\label{sec:grdm}

The differential diffuse $\gamma$--ray flux originating from WIMP
pair-annihilation, and arriving from a direction at an angle $\psi$
from the GC is given by~\cite{susy-dm-reviews}
\beq
\dfluxgde (\egamma, \psi) = \sum_{i} \frac{\sigma_i
v}{8\pi\mchi^2}\, \frac{d N^i_\gamma}{d\egamma}
\int_{\text{l.o.s.}} dl^{\prime}\, \rhochi^2(r(l^{\prime},\psi)),
\label{eq:diffgammaflux}
\eeq
where $\mchi$ denotes the WIMP mass and $\sigma_i v$ is a product of
the WIMP pair-annihilation cross section into a final state $i$ times
the pair's relative velocity and $d N^i_\gamma /d\egamma$ is the
differential $\gamma$--ray spectrum (including a branching ratio into
photons) following from the state $i$. Here we consider the total
contribution from both the continuum, resulting from cascade decays of
all kinematically allowed final state SM fermions and combinations of
gauge and Higgs bosons, and from photon lines coming from one loop
direct neutralino annihilation into $\gamma\gamma$ and $\gamma Z$. The
integral over the square of the DM mass density $\rhochi$ is taken
along the line of sight (l.o.s.) $l^{\prime}$ at an angle $\psi$
between us and the GC. It is related to the Galactic longitude ($l$)
and the Galactic latitude ($b$) by $\cos{\psi}=\cos{l}\cos{b}$.

It is
convenient to separate factors depending on particle physics and
on halo properties by introducing the dimensionless quantity~\cite{bub97}
\beq
J(\psi)\equiv 
\frac{1}{8.5\kpc}\left(\frac{1}{0.3\gev/\cmeter^3}\right)^2
\int_{\text{l.o.s.}} dl^{\prime}\, \rhochi^2(r(l^{\prime},\psi)).
\label{eq:jdef}
\eeq

The flux arriving from the angle $\psi$ is further averaged over the
solid angle $\Delta\Omega$ representing point spread function angle of
the detector or some wider angle of the sky, depending on the
situation, and one defines the quantity
\beq
{\jpsiave}_\deltaomega = \frac{1}{\deltaomega} \int_{\deltaomega}
d\Omega^{\prime} J(\psi^{\prime}).
\label{eq:javedef}
\eeq

The total flux from a solid angle $\deltaomega$ centered on the angle $\psi$ 
and integrated over photon energy $\egamma$ from an energy threshold
$E_{\text{th}}$ is then given by
\beq
\fluxg\deltaomega =  \int^{\mchi}_{E_{\text{th}}}
d\egamma\, \dfluxgdetext(\egamma, \Delta\Omega).
\label{eq:totalgrflux}
\eeq


One of the crucial ingredients in the analysis is the distribution of
dark matter in and around the Milky Way (MW). Unfortunately,
despite much effort and progress, this remains rather poorly known,
especially in the inner part of the MW. For this reason, in this
analysis we will consider some popular halo profiles with a strongly
varying cuspiness at small Galactic radius $r$.

Several popular profiles can be parametrized by~\cite{susy-dm-reviews}
\beq
\rhochi(r)= \rho_0
            \frac{\left[1+(r_0/r_{s})^\alpha\right]^{\frac{\beta-\gamma}{\alpha}}}
{ (r/r_0)^{\gamma}
            {\left[1+\left(r/r_{s}\right)^\alpha\right]^{\frac{\beta-\gamma}{\alpha}}}} ,
\label{eq:halomodels}
\eeq
where $\rho_0=0.3\gev/\cmeter^3$ represents the local DM
density. For example, in the Navarro, Frenk and White (NFW) model $r_0= 8.0\kpc$, $r_s= 20.0\kpc$,
$\alpha=1$, $\beta=3$ and $\gamma=1$~\cite{nfwhalo95}.
Another model that has recently become favored by numerical simulations of large structures is the 
Einasto model~\cite{einasto65,Navarro:2008kc} described by 
\begin{equation}\label{einasto}
\rho(r)=\rho_0\texttt{exp}[{\frac{2}{\alpha}
\frac{r_s^{\alpha}-r^{\alpha}}{r_{s}^{\alpha}}}], 
\end{equation}
where in this case $\rho_0= 0.066\gev/\cmeter^3$ and $r_{s}=25\kpc$, while the index
$\alpha$ takes the range
$\frac{1}{7}<\alpha<\frac{1}{3}$~\cite{Navarro:2003ew,Graham:2005xx},
with the best-fit value of $\alpha=0.17$~\cite{Diemand:2008in}. In
this paper, we will consider the best-fit case (called thereafter
``Einasto''), although one should remember that increasing $\alpha$
within the above range gives a whole range of profiles with decreasing
cuspiness.
The Burkert model~\cite{Burkert:1995yz} is an example of a realistic profile with a basically flat DM
density distribution in the inner region of the MW. It is parametrized by 
\beq \rhochi(r)= \rho_0 \frac{
  \left[1+(r_0/a_{n})\right]\left[1+(r_0/a_{n})^2\right] }{
  \left[1+(r/a_{n})\right]\left[1+(r/a_{n})^2\right] },
\label{eq:burkertmodel}
\eeq
where  $\rho_0= 0.34\gev/\cmeter^3$ and $a_{n}=11.68\kpc$.
\begin{figure}[tbh!]
\begin{center}
\includegraphics[width=8cm]{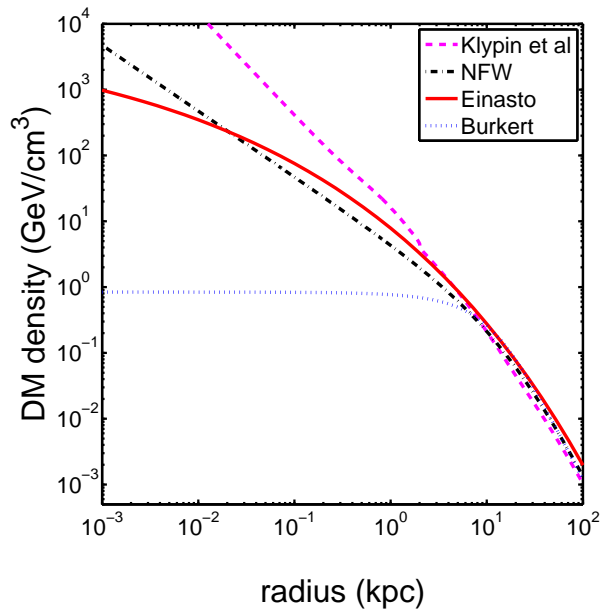}
\end{center}
\caption{Some popular dark matter halo profiles as a function of
  radius $r$. } \label{fig:halos}
\end{figure}
As the last profile, we consider the model of Klypin, \etal,~\cite{klypinmodel} as
providing the most divergent profile in the GC. It is based on the NFW
model but it is fitted to the data from the MW and takes into
account the effect of angular momentum exchange between baryons and
dark matter, and in this sense may be considered as more applicable to
our Galaxy than other profiles.  The inner radius density profile for the Klypin, \etal,
model is $\sim r^{-1.8}$ which is  steeper than the Einasto
and NFW profiles. Close to the solar radius all the models become quite
similar. As the Klypin, \etal, model results from the fits to the
data, no analytic expression is available. 

The above DM density profiles are presented in Fig.~\ref{fig:halos}.
We will use them as illustrative examples as they represent a large
variation in cuspiness at small $r$, which will be perhaps the single
most important property in determining DM contribution to diffuse
$\gamma$--radiation.



\section{Signatures of dark matter in  $\gamma$--ray spectrum from the
  Galactic Center}\label{sec:dmtestgc}

\begin{figure}[tbh!]
\begin{center}
\includegraphics[width=8cm]{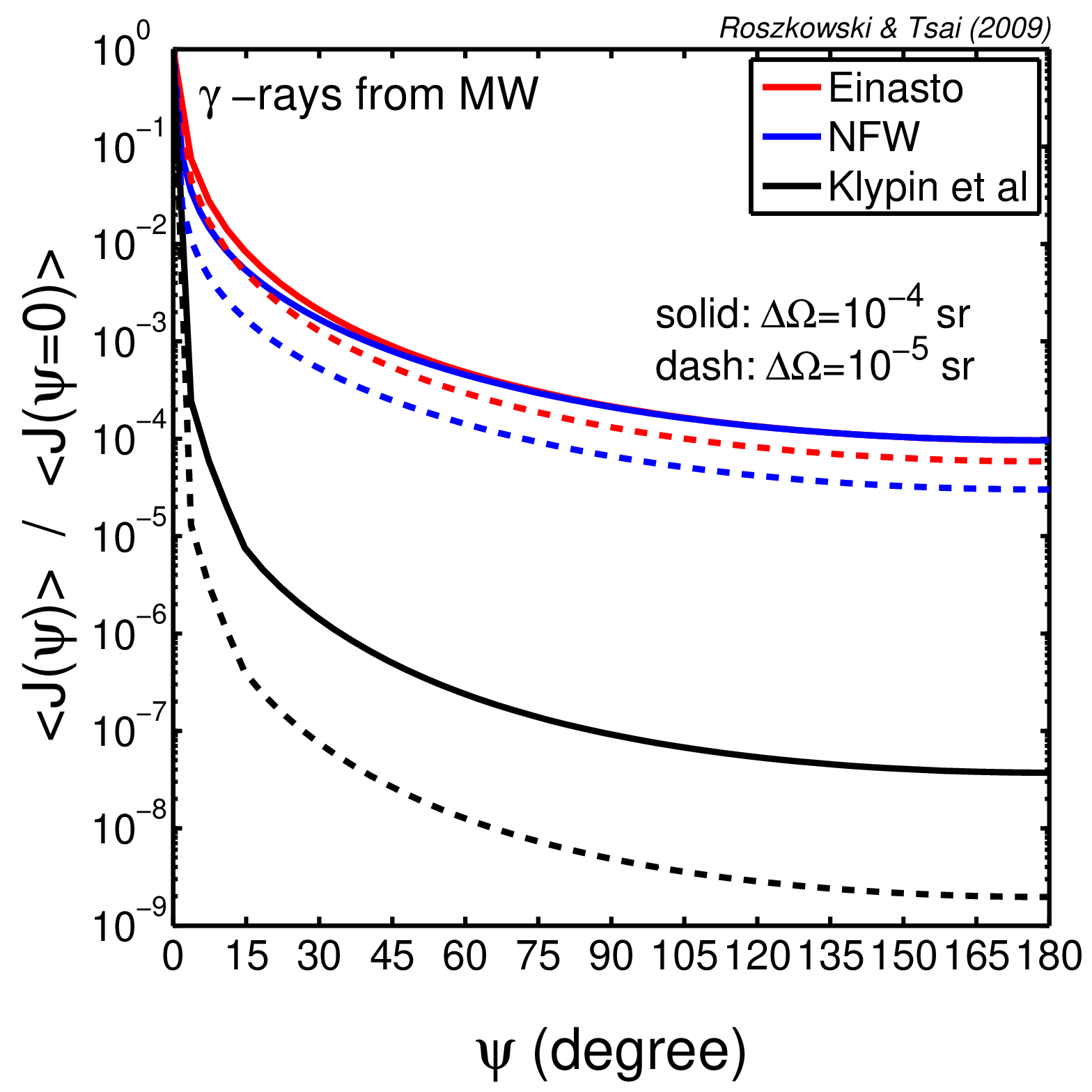}     
\end{center}
\caption{The ratio $\rdifffluxggc=
  \frac{{\jpsiave}_\deltaomega}{{\jpsizeroave}_\deltaomega} $ (the GC)
  versus $\psi$ for $\deltaomega=10^{-5}\sr$ (dashed) $10^{-4}\sr$
  (solid) for the halo models considered in the paper. The ratio does
  not depend on the DM particle mass nor its other microscopic
  properties. For the Galactic latitude $b=0$ the angle $\psi$
  coincides with the Galactic longitude, $\psi=\pm l$, while for $l=0$
  $\psi$ coincides with $\psi=\pm b$ (up to
  $90^{\circ}$). }\label{fig:cf_R_vs_angle}
\end{figure}

Ideally, one of the most convincing signatures of the annihilating DM
origin of the diffuse $\gamma$--radiation that Fermi's data could
produce, would be the flux with a spectrum that falls off with the
angle $\psi$ from the GC. This is because it is proportional to
$\rhochi^2$, with the DM density in most models decreasing with $r$,
or $\psi$, compare Eq.~(\ref{eq:diffgammaflux}) and
Fig.~\ref{fig:halos}.  Since the flux itself does depend on the DM
particle mass and its other unknown properties, as well as the photon energy
$\egamma$, we propose instead to consider the ratio
\beq \rdifffluxggc=
\frac{\dfluxgde(\egamma,\psi)}{\dfluxgde(\egamma,\psi=0)}=
\frac{{\jpsiave}_\deltaomega}{{\jpsizeroave}_\deltaomega}
=\frac{\int_{\text{l.o.s.}} dl^{\prime}\,
  \rhochi^2(r(l^{\prime},\psi))}{\int_{\text{l.o.s.}} dl^{\prime}\,
  \rhochi^2(r(l^{\prime},\psi=0))},
\label{eq:difffluxratio}
\eeq
which follows from Eqs.~(\ref{eq:diffgammaflux})--(\ref{eq:javedef})
and which is clearly  {\em only} dependent on the dark matter mass density
distribution in the halo. The square dependence of
$\rhochi$ along the l.o.s.  to a large degree cancels out, but the one in
the transverse direction does not, and provides a genuine effect of DM
annihilation in the region of the GC. The ratio~(\ref{eq:difffluxratio}) is shown in
Fig.~\ref{fig:cf_R_vs_angle} for three representative halo models and
for two different angular resolutions typical for Fermi LAT. (In the case of the more cuspy
profiles, in order to avoid a divergent behavior, we set a cutoff
radius of $r_c=10^{-5}\kpc$.)

Observing the ratio
$\rdifffluxggc$ 
shown in Fig.~\ref{fig:cf_R_vs_angle} would constitute a clear and
rather unambiguous signal of the DM origin of the diffuse
$\gamma$--radiation. Indeed, it is highly unlikely that any other,
known or unknown, source of diffuse $\gamma$--radiation would have a
distribution producing a similar fall-off. At least we are not aware
of any.\footnote{We stress that the behavior presented in in
  Fig.~\ref{fig:cf_R_vs_angle} applies to stable DM, whose pair
  annihilation is proportional to $\rhochi^2$. The flux from decaying
  DM instead depends on the DM density linearly and would produce a
  much flatter spectrum which would be probably harder to distantangle
  from astrophysical contributions.
} The astrophysical sources of the diffuse flux in the photon energy
range of interest are relatively well understood. The main contribution is
caused by primary cosmic ray (CR) interactions with interstellar
hydrogen and helium atoms, producing $\pi^0$'s which in turn decay via
$\pi^0\to\gamma\gamma$ with a peak at $m_{\pi}/2\simeq70\mev$, with a
symmetric distribution. Other sources include: inverse Compton fluxes
and bremsstrahlung due to CR scatterings of electrons and positrons
off the interstellar medium, isotropic background (\eg, extragalactic
radiation) and point sources. These contributions are well modelled
with GALPROP~\cite{galprop:ref} which in general reproduces current
data remarkably well.

Of course, as we said above, finding evidence for DM in the
measurement of the ratio ~(\ref{eq:difffluxratio}) would require a
highly cuspy profile at small $r$ and also relatively small
astrophysical background. While the former is poorly known and remains
a possibility, the latter is known to be hugely important
and is likely to overshadow, or even make it too difficult, to observe
the behavior shown in Fig.~\ref{fig:cf_R_vs_angle}. Firstly, the DM is
expected to be strongly enhanced only within the rather small range of
some $2-3^{\circ}$ around $\psi=0$. Secondly, there are many known (and
also likely many unresolved) point sources in the direction of the
GC. However, the DM signal should be basically spherical
while the astrophysical component is expected to be
disk-like.\footnote{We thank G.~J{\' o}hannesson for his comments on
  this point.} For this reason it will be of crucial importance to examine the
ratio shown in Fig.~\ref{fig:cf_R_vs_angle} in the special cases of
the Galactic plane ($b=0$) and the vertical plane $l=0$, where the
astrophysical component should show a different angular behavior but
the DM part should remain the same. The test is certainly going to be
very challenging but should be attempted as it would provide perhaps a
single most convincing test of the DM origin of the measured ratio.
Even if it is possible to measure the ratio
Eq.~(\ref{eq:difffluxratio}) only for some ranges of $\psi$, this
already could provide some vital information pointing towards the DM
origin of the effect.

Furthermore, if the measured ratio shows the behavior shown in
Fig.~\ref{fig:cf_R_vs_angle}, one could attempt to actually infer DM
profile, at least to some degree. In particular, if the ratio drops
rapidly at small $\psi$, this would imply a very cuspy profile, which
of course will also produce a much larger absolute flux at small $r$,
and for this reason will be much easier to detect.

Therefore, a measurement of the ratio Eq.~(\ref{eq:difffluxratio}), if
it confirms the behavior shown in Fig.~\ref{fig:cf_R_vs_angle}, will
provide a convincing signature of existence of DM in the region of the
GC, independently of the DM mass or its other properties, like
annihilation cross section or decay branching ratios of the
annihilation products. As a bonus, it may be possible to infer the
actual shape of the DM density profile.

A related important test of the DM origin would be a measurement of the
total $\gamma$--ray flux from the GC ($\psi=0$) as a function of the
total solid angle $\deltaomega$. Again, in order to remove  the
dependence on the WIMP properties, we introduce the ratio of the total
fluxes from the GC as a function of  $\deltaomega$,
\beq
\rfluxggc = \frac{\fluxg\deltaomega}{\fluxg{\deltaomega=10^{-5}\sr}},
\label{eq:totalfluxratio}
\eeq
where the choice
of $\deltaomega=10^{-5}\sr$ reflects Fermi LAT's angular resolution
but is otherwise fairly arbitrary. 

\begin{figure}[tbh!]
\begin{center}
\includegraphics[width=8cm]{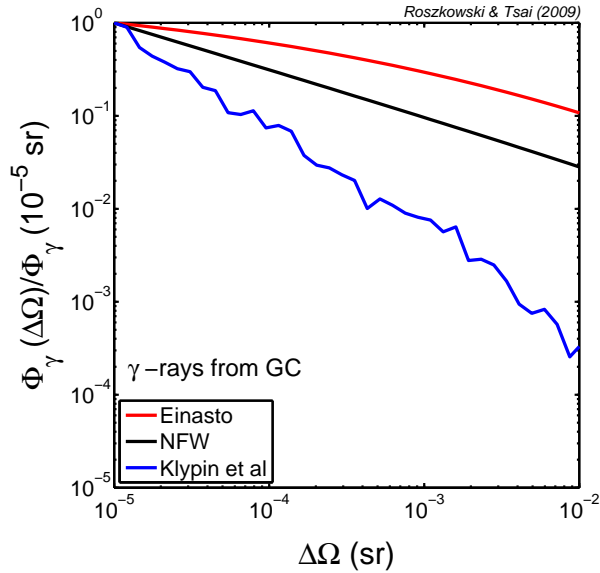}
\end{center}
\caption{The ratio of the total $\gamma$--ray flux $\fluxg\deltaomega$
  and $\fluxg{\deltaomega=10^{-5}\sr}$ from DM annihilation in the GC
  ($\psi=0$) versus $\deltaomega$ for three representative halo
  models. }\label{fig:GC_vs_delom}
\end{figure}

The quantity is shown in Fig.~\ref{fig:GC_vs_delom} for our three
representative halo models. (The curve for the Klypin, \etal, model
exhibits a non-smooth behavior because of numerical integration over
tabulated points, as there is no analytic expression available for the
profile.) Again, since the $\gamma$--ray flux from DM annihilation is
proportional to $\rhochi^2$, it would be hard to mimic the behavior
with conventional astrophysical sources. In addition, since the ratio
falls off more (less) rapidly for a more (less) cuspy DM profile, by
observing the approximate behavior of the ratio $\rfluxggc$ of the
measured total fluxes one could attempt to infer the cuspiness of the
DM halo profile in the area of the GC.  The quantities introduced in
Eqs.~(\ref{eq:difffluxratio}) and~(\ref{eq:totalfluxratio}) are not
independent as are both proportional, at some level, to the square of
DM halo density. Nevertheless, they are not the same as the latter
involves integration over the photon energy and also over the solid
angle $\deltaomega$ centered on the GC.

\section{Constraints on DM properties from Fermi LAT mid-latitude
  data}\label{sec:midlatconstr}

Next we move on to the discussion of the preliminary data on
mid-latitude $\gamma$-ray fluxes and the ensuing implications for DM
mass and distribution.  Last Spring, the Fermi Collaboration released
preliminary data on mid-lattitude $\gamma$-ray flux from the region of
the sky bounded by $10^{\circ}\leq |b|\leq 20^{\circ}$ and $0\leq l
<360^{\circ}$~\cite{porter_lodz09} (hereafter called $\deltaomega_{\rm
  mid-lattitude}$ for brevity), in the photon energy range between
$100\mev$ and $10\gev$. The spectrum revealed two important
features. Firstly, it did not confirm the spectrum measured over the
same area by EGRET, and is instead softer, with an integrated
intensity $J_{\rm LAT}(\egamma\geq 1\gev)=2.35\pm 0.01\times
10^{-6}\cmeter^{-2}\sec^{-1}\sr^{-1}$, compared to $J_{\rm
  EGRET}(\egamma\geq 1\gev)=3.16\pm 0.05\times
10^{-6}\cmeter^{-2}\sec^{-1}\sr^{-1}$, where only statistical errors
are shown and in both cases the contribution from point sources has
not been removed.

The second important feature of the Fermi LAT spectrum is that it
approximately agrees with the estimate of the contributions
astrophysical contributions, as computed by GALPROP (compare the right
panel of Fig.~1 of Ref.~\cite{porter_lodz09}) which, in considering DM
signatures, we will treat as background.

\begin{figure}[tbh!]
\begin{center}
\includegraphics[width=8cm]{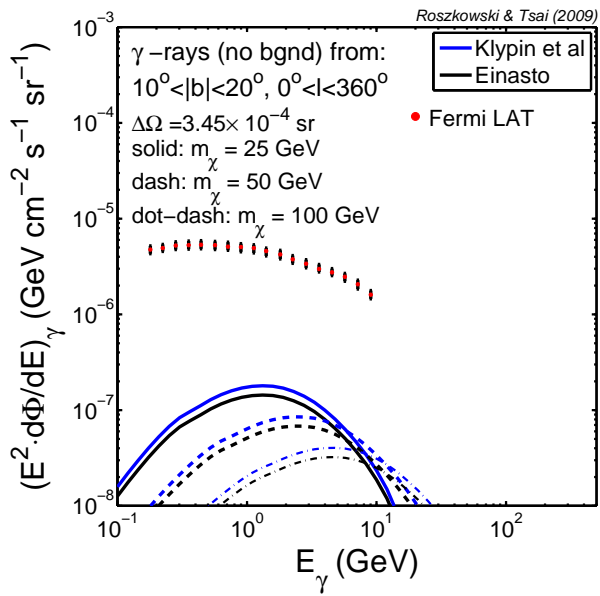}
\end{center}
\caption{The quantity $\egamma^2\dfluxgde$ of $\gamma$--ray flux
  from DM annihilation only averaged over $10^{\circ}\leq |b|\leq
  20^{\circ}$ and $0\leq l <360^{\circ}$ versus $\egamma$ for
  $\mchi=25\gev$ (solid), $50\gev$ (dash) and $100\gev$ (dot-dash) and
  the Klypin, \etal, model (blue) and the Einasto model with best-fit
  value of $\alpha=0.17$ (black). For comparison, the Fermi LAT data
  points are marked in red, and the associated $1\sigma$ errors in
  black. }\label{fig:b10_nobgnd}
\end{figure}

It is clear that the good agreement between the measured flux and the
astrophysical background puts some constraints on the allowed
contribution from DM annihilation. This is illustrated in
Fig.~\ref{fig:b10_nobgnd} where we plot the quantity $\egamma^2
\dfluxgdetext$ from DM pair annihilation 
computed using DarkSusy~\cite{Gondolo:2004sc} for a neutralino with mass $\mchi=25\gev$ (solid),
$50\gev$ (dash) and $100\gev$ (dot-dash) and the Klypin, \etal\ (blue)
or the Einasto (black) profiles, and averaged over $\deltaomega_{\rm
  mid-lattitude}$. In order to reproduce
$\sigmav=3\times10^{-26}\cmeter^3/\second$, typical of thermal WIMPs,
we chose the underlying parameters of the general Minimal Supersymmetric
Standard Model (MSSM) as follows: $\mu=-\mtwo=68.4\gev$,
$\mha=116.1\gev$ for $\mchi=25\gev$, $\mu=-\mtwo=113.3\gev$ and
$\mha=166.2\gev$ for $\mchi=50\gev$, and $\mu=211.9\gev$,
$\mtwo=-205.9\gev$ and $\mha=144.2\gev$ for $\mchi=100\gev$, in
addition to fixing $\tanb=10$, the average squark mass
$\msquark=2.5\tev$, 
$A_t=1.13\tev$ and $A_b=-1.92\tev$. 
For comparison, we show the Fermi LAT data. It is clear that, even for
rather small $\mchi$ and rather cuspy profiles, the DM contribution is
at least an order of magnitude too small. 

\begin{figure}[tbh!]
\begin{center}
\includegraphics[width=10cm]{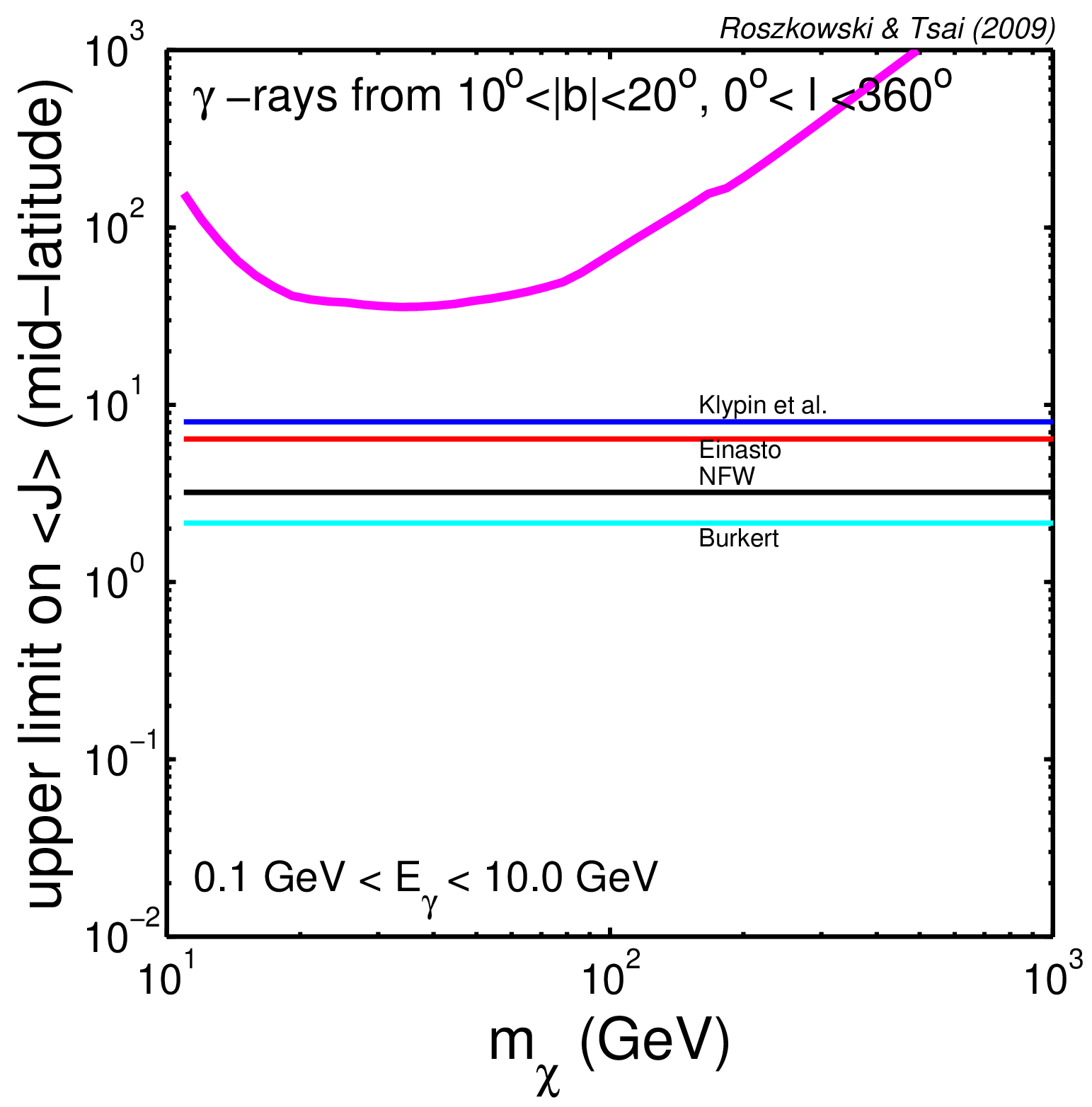}
\end{center}
\caption{An upper limit (solid magenta line) from mid-latitude ($10^{\circ}<|b|<20^{\circ}$
  and $0\leq l<360^{\circ}$) Fermi LAT data on $\javearea ({\rm
    mid-lattitude})$ versus the neutralino mass $\mchi$ taking the photon
  energy range of $0.1\gev<\egamma<10\gev$. For
  comparison, we show as horizontal lines $\jpsiave ({\rm
    mid-lattitude})$ for the Klypin, \etal, model (blue), the Einasto
  model with best-fit value of $\alpha=0.17$ (red), the NFW model
  (black) and the Burkert model (cyan), averaged over the same region of the sky. }\label{fig:javevsmchi}
\end{figure}

On the other hand, for small $\mchi$ the DM contribution would be too
big for much more cuspy profiles since it scales with $\rhochi^2$. This
allows us to derive an upper bound on such combinations.  This is
presented in Fig.~\ref{fig:javevsmchi} where on the horizontal axis we
plot the DM mass and on the vertical axis we plot the upper limit on
$\javearea ({\rm mid-lattitude})$, which is the quantity $\jpsi$ of
Eq.~(\ref{eq:jdef}) averaged over $10^{\circ}\leq |b|\leq 20^{\circ}$
and $0\leq l <360^{\circ}$. In deriving the upper limit we first establish
the amount that DM annihilations can contribute to the Fermi
LAT measurement of $\egamma^2\dfluxgde$ averaged over the mid-latitude
region $\deltaomega_{\rm mid-lattitude}$.
We do this by
reading out from the right panel of Fig.~1 of
Ref.~\cite{porter_lodz09} the Fermi LAT data and the astrophysical
background, and taking their difference. 
Next, for a given value of the neutralino mass $\mchi$ we use
Eq.~(\ref{eq:diffgammaflux}) to compute, up to a normalization
constant, $\javearea$ and ensure that it does not exceed the allowed
DM contribution. For definiteness we 
take the photon energy in the range $0.1\gev<\egamma<10\gev$, which is
the same as in the Fermi LAT data. We scan over the MSSM parameters
$\mu$, $\mtwo$ and $\mha$ in looking for the most conservative limit,
while ensuring that $\sigmav\simeq3\times10^{-26}\cmeter^3/\second$.
The spread in ${d N^i_\gamma}/{d\egamma}$ is only a few for
$\mchi\gsim50\gev$ but it increases into a few orders of magnitude at
lower $\mchi$, which explains the rise in the upper limit on
$\javearea$, shown in Fig.~\ref{fig:javevsmchi} as a solid magenta
curve. On the other hand, the DM contribution to the diffuse flux
drops down roughly as $1/\mchi^2$ and at large $\mchi$ the upper limit
again becomes much weaker. Varying $\egamma^2\dfluxgde$ of the Fermi
LAT data and of the background within their respective error bars
moves the upper limit on $\javearea ({\rm mid-lattitude})$ up and down
by a factor of a few but it still remains above the values $\javearea
({\rm mid-lattitude})$ for the Klypin, \etal, (blue), the Einasto
(red), the NFW (black) and the Burkert profile (cyan), which we have
plotted for comparison. Clearly, the upper limit, at present, puts a
rather mild, but non-trivial, constraint on the DM halo density cuspiness. 

It is clear from Fig.~\ref{fig:javevsmchi} that the Fermi LAT data
does put a constraint, albeit a rather mild one, on some combinations of DM mass and halo profile
at mid-latitudes. As more data is accumulated and analyzed from larger areas of the
sky, especially towards the Galactic Center, and assuming an adequate
agreement with astrophysical contributions, the upper limit on
$\javearea$ as a function of the DM particle mass may become
stronger. Nevertheless, it is interesting that, even with the current
data, one can already place some non-trivial constraints on the DM
halo profile as a function of its mass. Incoming data from the
region of the GC is likely to allow one to put a stronger limit on the
cuspiness, assuming of course no significant deviation from known astrophysical contributions.

\section{Summary}\label{sec:summary}

The hunt for a dark matter signal is in full swing, with several
different strategies and experiments reaching promising detection
sensitivities.  The LAT instrument on the Fermi Gamma-Ray Space
Telescope has already produced very high quality data on diffuse
$\gamma$--ray emission from intermediate Galactic latitudes and is
soon expected to provide a very high resolution and precision map
of the region of the Galactic Center where the density of dark matter
is expected to be enhanced. 

In this paper we have argued that, by examining an angular
distribution of the upcoming data from the GC, it may be possible to
see a contribution from DM annihilations for a cuspy enough DM
profile, and furthermore to get an estimate on the cuspiness,
independently of the mass and other properties of the WIMPs
constituting the DM. This could provide an unambiguous signature of the DM in the GC. 

Next, we showed that the recent Fermi data from mid-latitudes already
allows one to put an upper limit on the cuspiness of DM halo profile
as a function of the WIMP mass assumed to be the lightest neutralino
of minimal SUSY.  We are eagerly awaiting a release of a fuller set of
data from the GC and from the whole sky over the photon energies
extending to much larger values, $100\gev$ and beyond, which will
hopefully allow to shed more light on the properties of the dark
matter and its distribution in the Galactic halo.

While in deriving the upper bound in Fig.~\ref{fig:javevsmchi} we have
used the neutralino as a well-motivated case, this choice was not
essential and similar bounds could likely be derived for other stable
WIMP candidates.

\medskip {\bf Acknowledgments} \\ We are grateful to G.~Bertone, D.~Cumberbatch,
G.~J{\' o}hannesson and A.~Strong for valuable comments.



\end{document}